\documentclass[sigconf]{acmart}

\AtBeginDocument{%
  \providecommand\BibTeX{{%
    \normalfont B\kern-0.5em{\scshape i\kern-0.25em b}\kern-0.8em\TeX}}}

\setcopyright{acmcopyright}
\copyrightyear{2018}
\acmYear{2018}
\acmDOI{10.1145/1122445.1122456}

\acmConference[Woodstock '18]{Woodstock '18: ACM Symposium on Neural
  Gaze Detection}{June 03--05, 2018}{Woodstock, NY}
\acmBooktitle{Woodstock '18: ACM Symposium on Neural Gaze Detection,
  June 03--05, 2018, Woodstock, NY}
\acmPrice{15.00}
\acmISBN{978-1-4503-XXXX-X/18/06}

\begin{document}

\title{Multi-Modal Video Forensic Platform for Investigating Post-Terrorist Attack Scenarios}

\author{Alexander Schindler}
\affiliation{%
  \institution{Center for Digital Safety and Security \\ Austrian Institute of Technology}
  \city{Vienna}
  \country{Austria}
}
\email{alexander.schindler@ait.ac.at}

\author{Andrew Lindley}
\affiliation{%
  \institution{Center for Digital Safety and Security \\ Austrian Institute of Technology}
  \city{Vienna}
  \country{Austria}
}
\email{andrew.lindley@ait.ac.at}

\author{Anahid Jalali}
\affiliation{%
  \institution{Center for Digital Safety and Security \\ Austrian Institute of Technology}
  \city{Vienna}
  \country{Austria}
}
\email{anahid.jalali@ait.ac.at}

\author{Martin Boyer}
\affiliation{%
  \institution{Center for Digital Safety and Security \\ Austrian Institute of Technology}
  \city{Vienna}
  \country{Austria}
}
\email{martin.boyer@ait.ac.at}

\author{Sergiu Gordea}
\affiliation{%
  \institution{Center for Digital Safety and Security \\ Austrian Institute of Technology}
  \city{Vienna}
  \country{Austria}
}
\email{sergiu.gordea@ait.ac.at}

\author{Ross King}
\affiliation{%
  \institution{Center for Digital Safety and Security \\ Austrian Institute of Technology}
  \city{Vienna}
  \country{Austria}
}
\email{ross.king@ait.ac.at}

\renewcommand{\shortauthors}{Schindler, Lindley, Jalali, Boyer, Gordea, King}


\begin{abstract} 

The forensic investigation of a terrorist attack poses a significant challenge to the investigative authorities, as often several thousand hours of video footage must be viewed. Large scale Video Analytic Platforms (VAP) assist law enforcement agencies (LEA) in identifying suspects and securing evidence. Current platforms focus primarily on the integration of different computer vision  methods and thus are restricted to a single modality. We present a video analytic platform that integrates visual and audio analytic modules and fuses information from surveillance cameras and video uploads from eyewitnesses. Videos are analyzed according their acoustic and visual content. Specifically, Audio Event Detection is applied to index the content according to attack-specific acoustic concepts. Audio similarity search is utilized to identify similar video sequences recorded from different perspectives. Visual object detection and tracking are used to index the content according to relevant concepts. Innovative user-interface concepts are introduced to harness the full potential of the heterogeneous results of the analytical modules, allowing investigators to more quickly follow-up on leads and eyewitness reports.

\end{abstract}





\maketitle


\section{Introduction} 


The ability to promptly analyze mass video data with regard to its content is increasingly important for complex investigative procedures, especially for those dealing with crime scenes. Currently, this data is analyzed manually which requires hundreds or thousands of hours of investigative work. As a result, extraction of initial leads from videos after an attack takes a long time. Additionally, law enforcement agencies (LEA) may be unable to process all the videos, leaving important evidence and clues unnoticed. This effort further increases when evidence videos from civilian witnesses are uploaded multiple times from multiple sources. However, the prompt analysis of video data is fundamental – especially in the event of terrorist attacks – to prevent immediate, subsequent attacks.

A video analytic platform for large scale surveillance and forensic investigation has been reported in \cite{deangelusdemand}. This platform integrates a wider range of computer vision approaches to track persons or objects in CCTV video recordings. Similar platforms with a specific focus on facial recognition, identification and 3D reconstruction as well as spatial tracking have been reported in \cite{kolarow2013apfel,qu2018ivisx}. 
These platforms, studies on challenges for building large scale integrated video platforms \cite{mathew2017challenges} and corresponding test-beds \cite{gorodnichy2010vap} as well as surveys recommending future directions \cite{gorodnichy2016recognizing} only consider optimizing single-modal computer vision-based approaches.
The same applies to audio surveillance systems \cite{crocco2016audio} where considerable progress in detection accuracy is reported \cite{chandrakala2019environmental}. Literature on audio-visual or multi-modal integration of approaches in a large scale forensic or surveillance video analytic platforms is currently underrepresented.

\begin{figure*}[t] 
\centering
\includegraphics[width=1.0\textwidth]{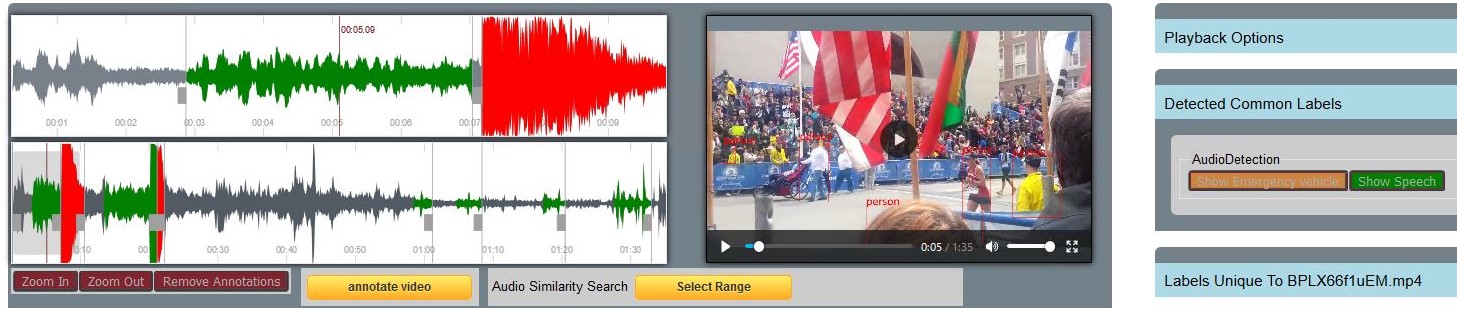}
\caption{Waveform representation for quickly analyzing content based on audio-visual events and acoustic similarity search}
\label{fig:SED}
\end{figure*}

%
In this paper we present a platform for forensic video analysis that fuses information from surveillance cameras and video uploads from eyewitnesses to assist law enforcement agencies (LEA) in identifying suspects and securing evidence by integrating analytical modules for different input modalities. The integration of multi-modal analytic modules represents a major contribution of this VAP. These modules can be categorized as indexing or search components. Indexing provides a fast entry point to an investigation through content-filters, according to pre-defined visual and acoustic events. Search components facilitate broad and extensive investigations by providing the means to find related video content based on a reference scene.
%
Accordingly, visualizations and user interfaces are introduced to efficiently harness the potential of the heterogeneous analytic results. These interfaces follow the following workflow: 1) filter the video content concerning specific events and objects such as provided by witnesses; 2) inspect filtered material and search for related videos to investigate the scene from different perspectives; 3) playback identified videos synchronized and simultaneously to identify suspects, evidence or new leads.



\begin{figure*}[t] 
\centering
\includegraphics[width=1.0\textwidth]{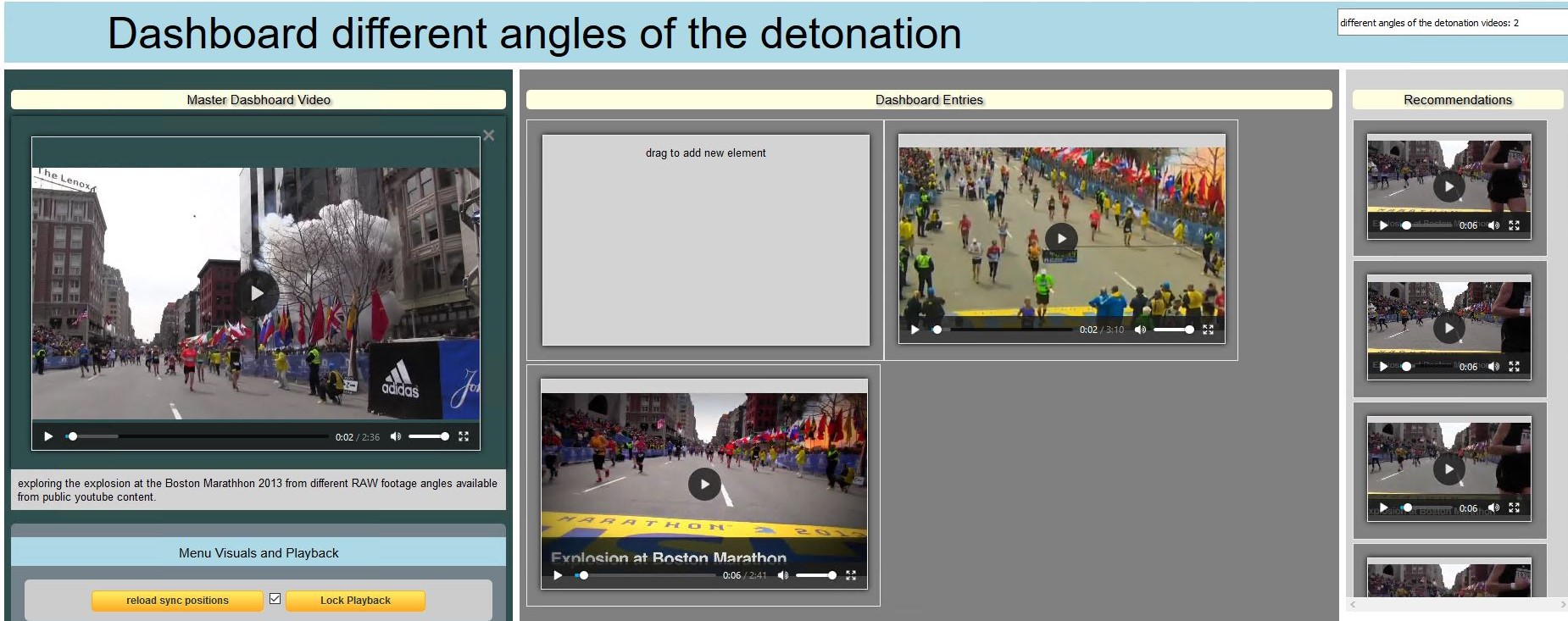}
\caption{User generated dashboard - different angles of an event based on acoustic fingerprints and video recommendations.}
\label{fig:SYNC}
\end{figure*}


\section{Video Forensic Platform}


The platform integrates analytical modules for different input modalities. The \textit{indexing modules} provide a fast entry into an observation by offering filters for attack-specific visual objects (e.g. persons, cars) or sound events (e.g. gunshots, explosions). Through these filters, case relevant videos are selected and ranked according the object prediction confidence, which ensures that the LEA agent starts the investigation with the most relevant videos. 

\subsection{Discrete Multi-Modal Video Indexing}

Witness reports often relate to discrete objects or events such as \textit{``there was a loud noise, then someone jumped into a blue car and drove away.''} The following modules provide methods to index the multi-modal content according to categories needed in a forensic crime-scene and post-attack scenario investigation.

\subsubsection{\textbf{Sound Event Detection (SED)}}

Requires indexing video files according scenario relevant sound events. The SED module facilitates the automatic temporal and categorical annotation of the events \textit{Gunshot, Explosion, Speech, Emergency vehicle} and \textit{Alarm}.

\vspace{-5pt}
\paragraph{Implementation:}

The implemented approach - detailed in \cite{schindler2019large} - is a combination of the models developed in the \textit{Detection and Classification of Acoustic Scenes and Events (DCASE)} \cite{mesaros2016tut} international evaluation campaign \cite{Lidy2016,schindler2016comparing,schindler2017multi} and the approach presented in \cite{xu2017attention}. The model applies a Convolutional Recurrent Neural Network (CRNN) \cite{schindler2019large} with an attention layer on log-scaled Mel-Spectrogram inputs (9.92 seconds audio, 44,1KHz sample rate, 80 Mel-bands, 2048 samples STFT-window size with 50\% overlap). It was trained on a pre-processed subset of the \textit{Audioset} dataset \cite{gemmeke2017audio}. Pre-processing included flattening of ontological hierarchies, resolving semantic overlaps, removing out-of-context classes (e.g. Music), re-grouping of classes and a final selection of task-relevant classes.

\vspace{-5pt}
\paragraph{Visualization:}

As the main entry point of the application, the AVP Search UI is divided into three areas: clipboard, filters and sorting and browsing of search results. It is the main entry point for content discovery and allows the user to combine labels from all audiovisual event detection algorithms and their required level of confidence in the queries. Other means for content discovery such as quick filters for textual metadata and user generated labels are also available. For each video it is possible to directly browse for acoustically similar videos or to add two videos to the clipboard for direct comparison. 

\subsubsection{\textbf{Visual Object Detection}}

Facilitates multi-target tracking - simultaneously detecting multiple targets at each time frame and matching their identities in different frames, yielding a set of target trajectories over time. 

\vspace{-5pt}
\paragraph{Implementation:}

The developed approach to a multi-class multi-target tracking method - also detailed in \cite{schindler2019large} - was trained and optimized on the specific scenario-relevant object categories. It is based on an appearance based tracker as in \cite{Wojke2017} and aims to add additional features such as targets motion and mutual interaction \cite{Sadeghian2017}, as well as learning temporal dependencies as in \cite{Ning2016}\cite{Sadeghian2017}. For each class, a multi-target tracker as in \cite{Wojke2017} is trained separately. During runtime, one tracker instance per class is activated, where all trackers are running in parallel. 

\vspace{-5pt}
\paragraph{Visualization:}

Detected objects by the video algorithms are directly displayed within the video playback using using color-coded bounding boxes and objects are traced between frames (see Figure \ref{fig:SED}). In addition, users are able to create custom annotations within the video which are also available as queries for content selection.

\subsection{Video Search and Retrieval}

Indexing videos according to predefined categories provides a fast way to start an investigation but it is limited by the type and number of classes defined, and undefined events such as \textit{train passing} cannot be detected. To overcome this obstacle and to facilitate the search for any acoustic pattern the following modules are provided.

\subsubsection{\textbf{Audio Similarity Estimation:}}

The audio similarity search serves to identify related segments. Based on the physical characteristics of sound as well as digital audio, videos with a similar acoustic texture must have been recorded in near vicinity. This facilitates spatio-temporal clustering and the identification of related video segments and consequently the investigation of the scene from different viewpoints.

\vspace{-5pt}
\paragraph{\textbf{Implementation:}}

The approach for estimating audio similarity is based on \cite{schindler2016europeana} and applies a broad rage of weighted audio features \cite{schindler2019large}. Features are extracted for each six seconds of audio content of every video file and the vector distances are calculated between all these features, enabling a sub-segment similarity search. 

\vspace{-5pt}
\paragraph{Visualization:}

The AVP provides a dedicated UI for acoustic event detection and analysis. SEDs and their aggregated timespans are color-coded within an audio waveform representation that is synchronized with the video. This allows users to quickly play back relevant events, highlight multiple sound event detections and to directly jump between detected timeframes. Based on a selected master SED and timestamp, the UI allows users to easily query for acoustically similar videos based on the detected event and to synchronously playback the similar scenes over all videos.

\subsubsection{\textbf{Audio Synchronization}}

The audio synchronization module temporarily aligns video pairs according to their acoustic texture, because the temporal video-metadata is not regarded as reliable. This alignment acts as basis for temporal filtering and is relevant for the simultaneous playback of videos.

\vspace{-5pt}
\paragraph{Implementation:}

The technique used for audio synchronization is based on the work of Kennedy and Naaman \cite{kennedy2009less}, where a two-phase approach is implemented to sync the audio data: a fingerprinting phase and a matching phase. To fingerprint the wave file, first we form a neighborhood of morphological constructs and detect the local frequency peaks within each neighbourhood. Afterwards, we calculate the distance between each two peaks and their exact location in the audio spectrogram to create an unambiguous hash value, as well as storing their offsets. These two combinations identify the audio landmark fingerprints. To synchronize two audios, first we need to make sure if the two recordings are matching files (contain the same audio data with a shift in their start point). To do so, we group the offsets of matched hashes into 50ms bins and normalize their counts to compute a threshold based on their mean and standard deviation. A significant large number of bins exceeding the threshold indicates that the audios are a match. And their offsets is used to determine the shift between the two audio files. 
This approach has proven to accurately identifying the matched pairs as well as a fast computation of the fingerprints. As noted in \cite{kennedy2009less}, factors influencing on the success of this method strongly depends on the audio quality and audio length. This was also observed in our experiments, where we attempted to synchronize audio files of post terrorist attacks and longer tracks provided more information for extracting fingerprints. 

\vspace{-5pt}
\paragraph{Visualization:}
The AVP provides a dedicated audio synchronisation dashboard, which for example is able to synchronize playback of the same event from different angles. Based on a selected master video, a user selects a audio synchronization point with relevance to the created dashboard, for example an explosion. Based on these fingerprints recommendations of similar video segments are presented. The overall accuracy of video recommendations increases the more videos a user adds to the dashboard.  


\section{Demonstrator}

The Audio and Video Analytics prototype (AVP) is a standalone solution for the visualization and interaction with key components developed in the areas of audio event and video object detection. It provides a user interface to support the use cases (1) content detection (2) video similarity analysis (3) video synchronization based on acoustic events and (4) user generated video annotation.

\vspace{-5pt}
\paragraph{Implementation:}

A local instance requires a Java Runtime Environment and ships with out of the box configurations for the Tomcat 7 webserver and PostgreSQL 11 database. The AVP consists of (1) HttpServlets REST-API and (2) server-side business logic and configuration management implemented in Java (Data Access Objects, Jackson databinding, JPA persistence) as well as (2) static HTML templates. The latter include placeholders for expected data, JavaScript for on-page interaction and corresponding HTMLBuilder classes as controllers for populating the pages on the fly for the use cases similarity analysis, sync analysis, dashboard and search. These are expanded by Ajax full-text and paging as well as a media streaming servlet for the delivery of the local media content to support for example drag and drop for easy content selection in the application. Maven is used as the build environment, which defines the module dependencies and the packaging of the web application. The final demonstrator contains over 142 GB of video data and the associated detection results.

\vspace{-5pt}
\paragraph{Customizability:}
A loose coupling between the UI and the application's content is achieved by relying on standardized schemas for integrating algorithmic event detection results. When the application is called, an initial database schema and index to support search is generated based on all available JSON event detection artefacts. This includes user-generated annotations and dashboards, as they follow the same process for extended portability. The reliance on the actual JSON artefacts in the interaction and visualisation of information within the application allows a developer to easily extend the prototype with different levels of detection algorithms (e.g. initial quick analysis vs. longer running in-depth analysis) and provide results once they are available.

\begin{figure}[t] 
\centering
\includegraphics[width=1.0\columnwidth]{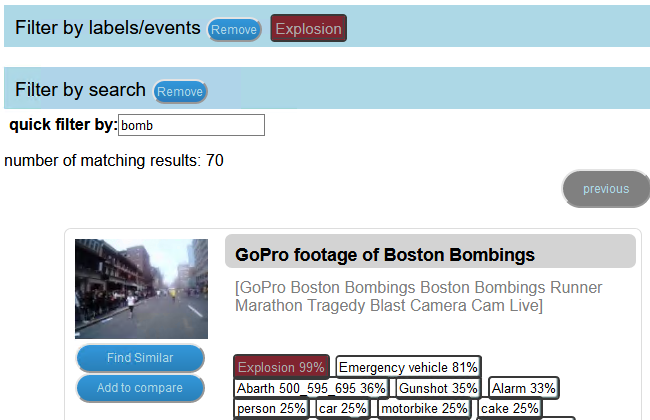}
\caption{content discovery by combining events from the audio visual detection algorithms and required minimum level of confidence of detection results}
\label{fig:SEARCH}
\end{figure}

\section{Conclusions and Future Work} 

The described platform integrates audio-visual analysis modules and introduces visualizations and user interfaces to harness this heterogeneous information intuitively. This in turn enables a quick start to and rapid progress in investigations, with the audio event detection module serving as a fast entry point. Based on these events, investigators can navigate through the video content, either by using the visual tracking modules on identified persons or objects, or by using the audio similarity search to find related video content, thereby increasing the chances for identification.

Future work will focus on integrating text as a further modality to harness information provided by social media platforms. We will also extend the audio similarity search to adaptively learned and optimized audio representations \cite{schindler2019multi}.


\begin{acks}
This work was supported by the FFG KIRAS project \textit{Flexible, semi-automatic Analysis System for the Evaluation of Mass Video Data (FLORIDA)} and was further developed under the EU Horizon 2020 project \textit{VICTORIA} / SEC-740754.
\end{acks}

\bibliographystyle{ACM-Reference-Format}
\bibliography{references}

\end{document}